\documentclass[twocolumn]{aastex631} 
\usepackage{amsmath} 

\begin{document}

\title{Early-time millimeter observations of the nearby Type II SN~2024ggi}

\author[0000-0003-3031-6105]{Maokai Hu}
\affiliation{Physics Department, Tsinghua University, Beijing 100084, China}
\email{kaihukaihu123@mail.tsinghua.edu.cn}
\author[0000-0003-3139-2724]{Yiping Ao}
\affiliation{Purple Mountain Observatory, Chinese Academy of Sciences, Nanjing 210023, China}
\affiliation{School of Astronomy and Space Science, University of Science and Technology of China, Hefei 230026, China}
\author[0000-0002-6535-8500]{Yi Yang}
\affiliation{Physics Department, Tsinghua University, Beijing 100084, China} 
\author[0000-0001-7201-1938]{Lei Hu}
\affiliation{McWilliams Center for Cosmology, Department of Physics, Carnegie Mellon University, Pittsburgh, PA, USA}
\author[0000-0002-5323-2302]{Fulin Li}  
\affiliation{Purple Mountain Observatory, Chinese Academy of Sciences, Nanjing 210023, China}
\author[0000-0001-7092-9374]{Lifan Wang}
\affiliation{George P. and Cynthia Woods Mitchell Institute for Fundamental Physics \& Astronomy, Texas A. \& M. University, 4242 TAMU, College Station, TX 77843, USA}
\email{lifan@tamu.edu}
\author[0000-0002-7334-2357]{Xiaofeng Wang}
\affiliation{Physics Department, Tsinghua University, Beijing 100084, China} 
\email{wang\_xf@mail.tsinghua.edu.cn}





\begin{abstract}

The short-lived ionized emission lines in early spectroscopy of the nearby type II supernova SN~2024ggi signify the presence of dense circumstellar matter (CSM) close to its progenitor star.
We proposed the Atacama Large Millimeter/submillimeter Array (ALMA) observations by its Director's Discretionary Time program to catch the potential synchrotron radiation associated with the ejecta$-$CSM interaction. Multi-epoch observations were conducted using ALMA band 6 at +8, +13, and +17 days after the discovery. The data show non-detections at the position of SN~2024ggi with a $3\sigma$ upper limit of less than 0.15 mJy, corresponding to a luminosity of approximately $8\times10^{24}\ {\rm erg}\,{\rm s}^{-1}\,{\rm Hz}^{-1}$. In this paper, we leverage the non-detections to place constraints on the properties of CSM surrounding SN~2024ggi. We investigate both the Wind and Eruptive models for the radial distribution of CSM, assuming a constant mass-loss rate in the Wind model and a distance-variant mass-loss rate in the Eruptive model. The derived CSM distribution for the Wind model does not align with the early-time spectral features, while the ALMA observations suggest a mass-loss rate of $\sim 5\times10^{-3}\ {\rm M}_{\odot}\,{\rm yr}^{-1}$ for the Eruptive model. Conducting multi-epoch millimeter/submillimeter observations shortly after the explosion, with a cadence of a few days, could offer a promising opportunity to capture the observable signature of the Eruptive model. 

\end{abstract}



\section{Introduction} \label{sec:intro}

Core-collapse supernovae (SNe) result from the explosive demise of massive stars at the end of their nuclear burning, and these explosive events carry out critical information about progenitor stars (e.g., \citealt{2012ARA&A..50..107L}). An evident product of the pre-explosion evolution is the circumstellar matter (CSM), generated from the mass-loss process of progenitors like the stellar wind, the accretion process, or the eruptive outburst (e.g., \citealt{2024ApJ...974..270C,2024ApJ...963..105M}). The interaction between the SN ejecta and pre-existing CSM leads to various observable phenomena from X-ray (thermal bremsstrahlung emission, \citealt{2012ApJ...747L..17C,2012ApJ...759..108S,2022ApJ...928..122M}) and optical (thermal black-body emission, \citealt{1994ApJ...420..268C,Wood-Vasey:2004ApJ...616..339W,2013MNRAS.435.1520M}) to radio (synchrotron radiation, \citealt{1982ApJ...259..302C,1998ApJ...499..810C,2022ApJ...934....5Y}) bands. Therefore, the multi-band signals relating to the existence of CSM can trace the mass-loss history of massive stars before their death.  

\begin{figure*}[t]
\centering
\includegraphics[width = 0.95 \textwidth]{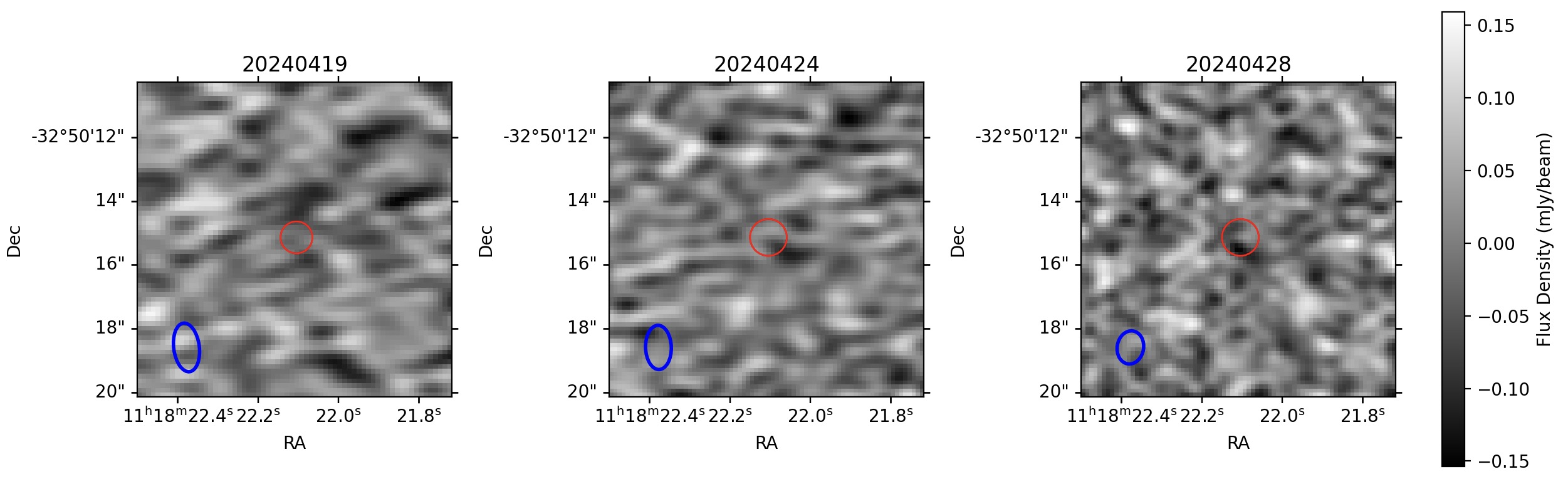}
\caption{Continuum images of SN~2024ggi at 1.3mm taken with ALMA at three epochs. The red circles, each with a diameter of 1 arcsec, indicate the location of SN~2024ggi at the coordinate RA/DEC = 11:18:22.087, -32:50:15.27. A color bar on the right shows the value scaling in units of mJy/beam. The synthesized beams are shown at the bottom-left of each plot.} 
\label{fig_00} 
\end{figure*} 

In particular, radio emission plays an essential role as an indicator of the CSM interaction. Type IIn SNe, characterized by narrow emission lines in their spectra, have been pursued for decades by radio facilities. These objects tend to fade slowly due to the enduring interaction with extended CSM and hence show signals in radio radiation for years since the discovery (e.g., \citealt{1996AJ....111.1271V,2009ApJ...690.1839C,2012ApJ...755..110C,2015ApJ...810...32C,2021ApJ...908...75B,2022ApJ...938...84D,2024A&A...686A.129S}). In contrast, the dense CSM close to the progenitor star will be rapidly swept up by the SN ejecta and then transit into collisionless shock. This dynamic process could introduce fast-evolving optical signatures (e.g., \citealt{2014Natur.509..471G,2016ApJ...818....3K,2017NatPh..13..510Y,2020MNRAS.498...84Z,2021MNRAS.505.4890L,2024ApJ...965...85A,2023arXiv231114409L}). Meanwhile, the SN explosion surrounded by confined CSM may also produce fast-evolving radiation in radio bands, especially in millimeter bands, as the synchrotron self-absorption and free-free absorption are severe in low frequencies. 

SN~2023ixf, which exploded in the nearby galaxy M101, is a type II SN with short-lived ionized emission lines, indicating the existence of confined dense CSM \citep{2023ApJ...956L...5B,2023ApJ...955L...8H,2023ApJ...954L..42J,2023ApJ...956...46S,2023SciBu..68.2548Z}. The corresponding mass-loss rate of the CSM surrounding SN~2023ixf is about $10^{-2}\ {\rm M}_{\odot}\,{\rm yr}^{-1}$, inferred from fitting the optical light curve or the emission line \citep{2023arXiv231114409L,2023arXiv231010727Z}. The millimeter-band observation of SN~2023ixf with the Submillimeter Array has yielded a consistent result, placing constraints on the mass-loss rate either larger than $10^{-2}\ {\rm M}_{\odot}\,{\rm yr}^{-1}$ or less than $10^{-6}\ {\rm M}_{\odot}\,{\rm yr}^{-1}$ \citep{2023ApJ...951L..31B}. 

SN~2024ggi, another nearby type II SN with early-time ionized emission lines, exploded one year later, providing a rare opportunity to catch its millimeter signals relating to the CSM interaction. This paper presents the millimeter-band observation of SN~2024ggi with the Atacama Large Millimeter/Submillimeter Array (ALMA) through a Director's Discretionary Time program (PI: M. Hu). Section~\ref{SecII} provides details of the observation and data. The adopted CSM model is described in Section~\ref{SecIII}, and Section~\ref{SecIV} outlines the result. Section~\ref{SecV} shows the discussion and conclusion. 

\section{Observations and Data}
\label{SecII}

\begin{figure*}[t]
\centering
\includegraphics[width = 0.95 \textwidth]{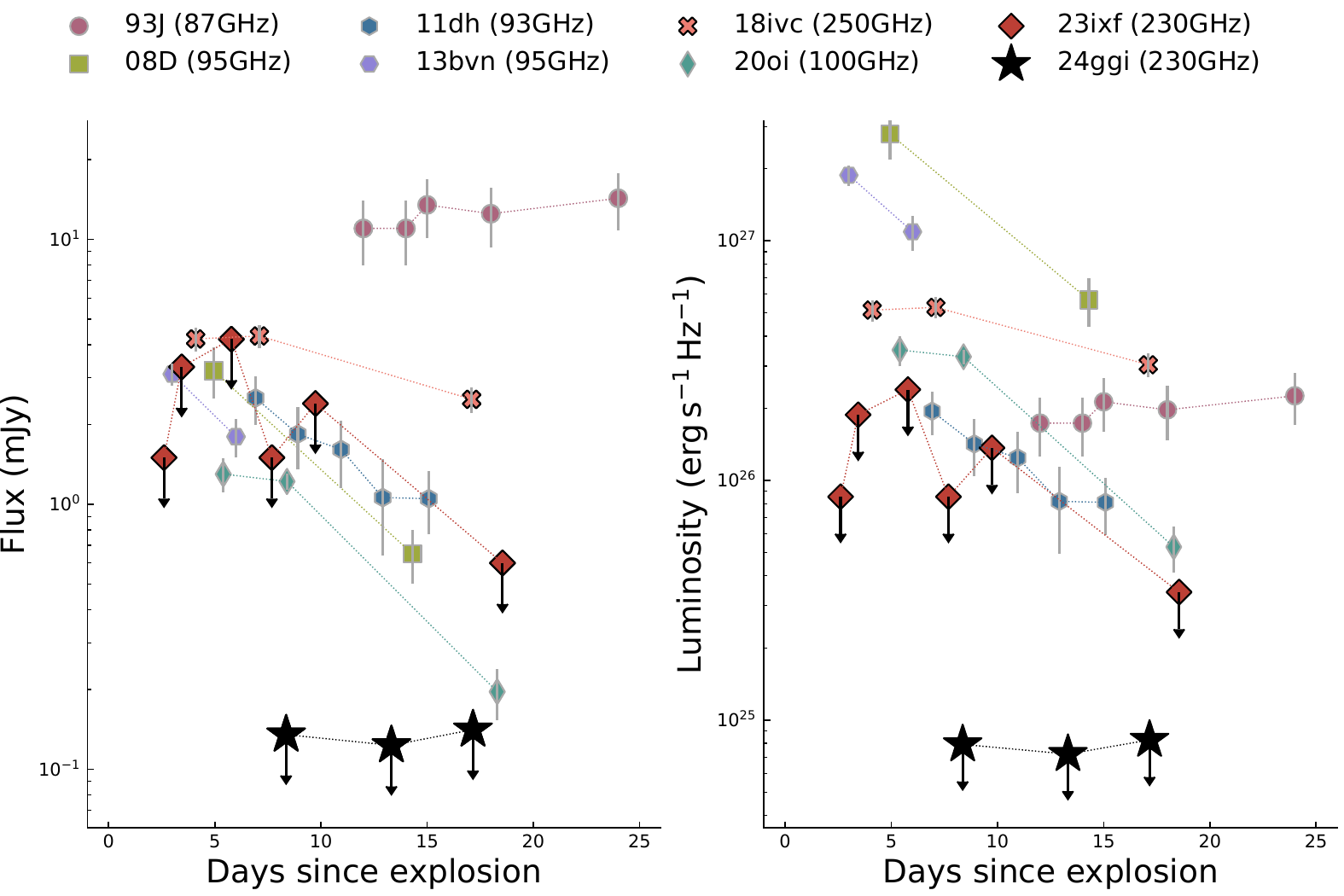}
\caption{Left panel: the early-time millimeter-band observations of core-collapse SNe, including SN~1993J (\citealt{1993IAUC.5775....1P,2007ApJ...671.1959W}), SN~2008D (\citealt{2008Natur.453..469S,2010A&A...522A..14G}), SN~2011dh (\citealt{2013MNRAS.436.1258H}), iPTF13bvn (\citealt{2013ApJ...775L...7C}), SN~2018ivc (\citealt{2023ApJ...942...17M}), SN~2020oi (\citealt{2021ApJ...918...34M}), SN~2023ixf (upper limits, \citealt{2023ApJ...951L..31B}), and SN~2024ggi ($3\sigma$ upper limits). The corresponding luminosity is shown in the right panel.} 
\label{fig_11} 
\end{figure*} 

SN~2024ggi was discovered by the Asteroid Terrestrial-impact Last Alert System on 2024 April 11.14 UT in the nearby galaxy UGC200 ($\sim$ 7.0 Mpc) and then classified as type II SN with early-time ionized emission lines \citep{2024arXiv240609270C,2024TNSTR1020....1T,2024TNSAN.104....1Z}. The progenitor star is possibly a red supergiant from the pre-explosion archival images \citep{2024TNSAN.100....1S,2024ApJ...969L..15X}. The follow-up observations show strong evidence of fast-evolving ionized emission lines and rapid brightening, indicating the interaction between SN ejecta and confined dense CSM \citep{2024A&A...688L..28P,2024ApJ...972L..15S,2024ApJ...970L..18Z}. Both photometric and spectroscopic observations suggest the discovery date is close to the explosion of SN~2024ggi. We adopted the explosion time from \cite{2024A&A...688L..28P} when we model the millimeter-band data of SN~2024ggi in the subsequent section.

\begin{table}[h]
    \centering
    \begin{tabular}{c|ccc}
    \hline
    Observation date  &   2024-04-19  & 2024-04-24 &  2024-04-28 \\ 
    \hline
    Phase & 8.4 days  &  13.3 days  &  17.2 days  \\
    Antennas number   &   46   &  44  &  43 \\
    Integration time  &   300 s  &  300 s  &  300 s \\
    Major axis        &   1.5$^{\prime\prime}$   &  1.2$^{\prime\prime}$   & 0.91$^{\prime\prime}$  \\
    Minor axis        &   0.8$^{\prime\prime}$   &  0.7$^{\prime\prime}$ & 0.72$^{\prime\prime}$ \\
    Position angle     &   -82$^{\circ}$  & -88$^{\circ}$  &   81$^{\circ}$ \\
    Image RMS         &  0.045 mJy   &  0.041 mJy &  0.047 mJy \\
    $3\sigma$ upper limits &  0.136 mJy &  0.124 mJy &  0.141 mJy\\
    \hline
    \end{tabular}
    \caption{The ALMA observation of SN~2024ggi (230 GHz). The phase corresponds to the explosion time from \cite{2024A&A...688L..28P}. The 3$\sigma$ upper limits are estimated assuming the source is unresolved by ALMA.}
    \label{tab:my_label}
\end{table}

SN~2024ggi was observed using ALMA, centered at the coordinate RA/DEC = 11:18:22.087, -32:50:15.27, over three epochs. The observations were conducted in continuum mode with a total bandwidth of 7.5 GHz in band 6, centered at a frequency of 223 GHz. J1126-3828 was used as the phase/gain calibrator for all three epochs, while J1427-4206, J1107-4449, and J1037-2934 served as flux calibrators, each on a different epoch. Data was processed using the Common Astronomy Software Application (CASA) pipeline, version 2023.1.0.124. The primary beam size at the central frequency of 223 GHz is approximately 26 arcsec. The full-width at half-maximum of the achieved synthesized beam size ranges from 0.8 to 1.1 arcsec. The continuum maps have a root-mean-square (RMS) noise level of around 0.041-0.047 mJy/beam. Detailed information about the observations is provided in Table~\ref{tab:my_label}. Figure~\ref{fig_00} displays the continuum maps towards the target, showing no detected emission from SN~2024ggi across the three epochs in the millimeter band. Unfortunately, bad weather conditions prevented us from conducting submillimeter-band observations. 

In this paper, we present the early-time observations by ALMA spanning about +8 -- +17 days after the discovery. The three ALMA observations were conducted at Band 6 to cover the frequency of 223 GHz, corresponding to the 230 GHz at the rest frame. Figure~\ref{fig_11} compares the 3$\sigma$ RMS sensitivity of SN~2024ggi and the millimeter-band observation of other core-collapse SNe. Thanks to ALMA's high sensitivity and high spatial resolution, the obtained image sensitivity and the corresponding luminosity of SN~2024ggi is about one magnitude deeper than that of SN~2023ixf. This improvement allows us to impose strict constraints on the circumstellar environment surrounding SN~2024ggi.

\section{Methods} 
\label{SecIII}

\begin{figure*}[t!]
\centering
\includegraphics[width = 0.95 \textwidth]{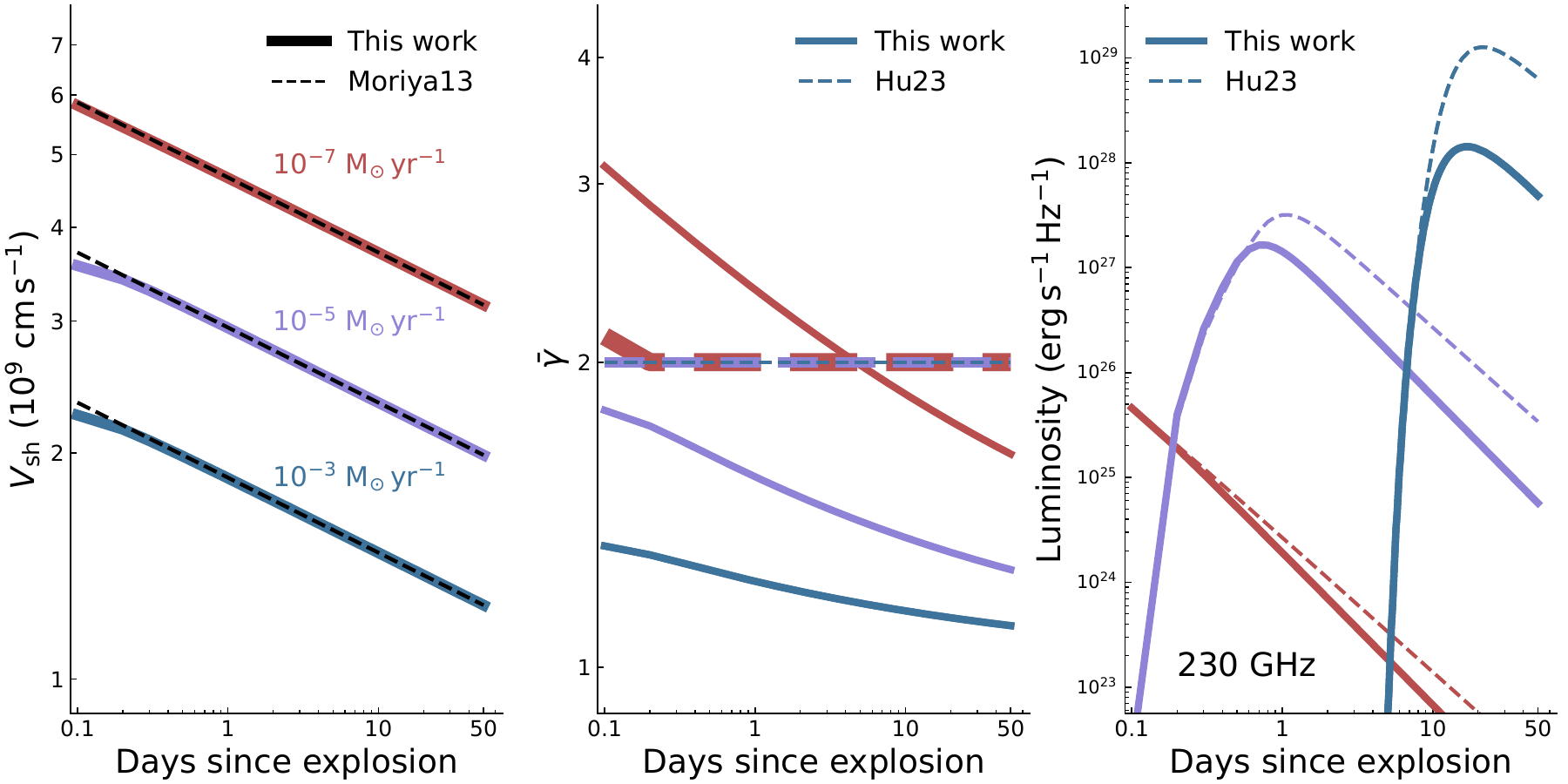}
\caption{In the context of the Wind model, this figure displays the performance comparison for the shock velocity ($V_{\rm sh}$, left panel), the mean $\gamma$ ($\bar{\gamma}$, middle panel), and predicted luminosity at 230 GHz (right panel)  with three sets of mass-loss rate, $\dot{M}_{\rm w} = 10^{-7}\ {\rm M}_{\odot}\,{\rm yr}^{-1}$ (red lines), $10^{-5}\ {\rm M}_{\odot}\,{\rm yr}^{-1}$ (purple lines), and $10^{-3}\ {\rm M}_{\odot}\,{\rm yr}^{-1}$ (blue lines), respectively. The left panel is the comparison of $V_{\rm sh}$ between our numerical code (solid lines, labeled as 'This work') and the formula in \cite{2013MNRAS.435.1520M} (dashed black lines, labeled as 'Moriya13'), confirming the validity of our model. The middle and right panels display the performance comparison between the calculation adopted in this work (solid lines, labeled as 'This work') and in \cite{2023MNRAS.525..246H} (dashed lines, labeled as 'Hu23') for $\bar{\gamma}$, and the predicted luminosity, respectively. The ejecta energy is $2\times10^{51}\ {\rm erg}$ for all calculations in this figure. } 
\label{fig_22} 
\end{figure*} 

\subsection{Radial distribution of CSM}
\label{Sec301}

We adopted the code developed by \cite{2023MNRAS.525..246H} to calculate the dynamical process when the ejecta crashes into CSM and predict the associated synchrotron radiation. A typical scenario of the radial distribution of CSM is the Wind model with CSM density ($\rho_{\rm csm}$) as $\rho_{\rm csm} = \dot{M}_{\rm w}/4\pi v_{\rm w}R^2$, where $v_{\rm w}$ is the wind velocity, $R$ is the distance to SN, and $\dot{M}_{\rm w}$ is the mass-loss rate of CSM. In the Wind model, $\dot{M}_{\rm w}$ is assumed to be constant, and the shock velocity ($V_{\rm sh}$) can be explicitly expressed as the function of the ejecta energy ($E_{\rm ej}$), the ejecta mass ($M_{\rm ej}$), $\dot{M}_{\rm w}$, and the time since the explosion ($t$) (e.g., \citealt{2013MNRAS.435.1520M}). As shown in the left panel of Figure~\ref{fig_22}, $V_{\rm sh}$ calculated from our numerical code is consistent with the formula from \cite{2013MNRAS.435.1520M}, suggesting the validity of our numerical code. 

On the other hand, the progenitor star may experience an eruptive mass-loss process before the SN explosion. \cite{Hu_inprep} employed a piece-wise function to describe the radial distribution of $\dot{M}_{\rm w} (R)$ relating to an Eruptive model as below, 
\begin{scriptsize}
\begin{align} 
\label{eq_Mw}
\dot{M}_{\text{w}}(R) = \begin{cases} 
0, & R \le R_0 \\
\dot{M}_{\rm w,0}(\frac{R - R_0}{R_1 - R_0})^{n_1}, & R_0 < R \le R_1 \\ 
\dot{M}_{\rm w,0}, & R_1 < R \le R_2 \\
(\dot{M}_{\rm w,0} - \dot{M}_{\rm w, min}) (\frac{R_3 - R}{R_3 - R_2})^{n_2} + \dot{M}_{\rm w,min}, & R_2 < R \le R_3 \\
\dot{M}_{\rm w, min}, & R > R_3\\
\end{cases}
\end{align} 
\end{scriptsize}
where $\dot{M}_{\rm w,0}$ is a characteristic value of mass-loss rate associated with the eruptive process, and $\dot{M}_{\rm w,min}$ represents the stellar wind before the eruption. 
The parameters $R_0$, $R_1$, $R_2$, and $R_3$ describe the distance-variant mass-loss rate of the pre-existing CSM, that $\dot{M}_{\rm w}(R)$ increases to $\dot{M}_{\rm w,0}$ within the distance of $R_1$ and decreases to $\dot{M}_{\rm w,min}$ from $R_2$ to $R_3$. 

The Eruptive model described by the above piece-wise function mainly includes two CSM regions, that are the inner one with high mass-loss rate of $\dot{M}_{\rm w,0}$ and the outer one with low mass-loss rate of $\dot{M}_{\rm w,min}$. This feature of our Eruptive model is consistent with other studies (e.g., \citealt{2021ApJ...909..209P,2023arXiv231010727Z,2024arXiv241017580C}), although some differences exist. This simplified Eruptive model is also consistent with the hydrodynamic simulation \citep{2024ApJ...974..270C}. For the Eruptive model, the dynamical process of the ejecta$-$CSM interaction can be numerically resolved using the code developed in \cite{2023MNRAS.525..246H}. Note that neither the Wind model nor the Eruptive model in our study considers the binary system scenario, so the geometric distribution of pre-existing CSM is spherical. 

\subsection{Synchrotron radiation}

The theory and the technical method of synchrotron radiation generated from the ejecta$-$CSM interaction has been well established \citep{1982ApJ...259..302C,1998ApJ...499..810C,2014ApJ...787..143B,2014ApJ...792...38P,2020ApJ...890..159L}. In this paper, we will update the synchrotron radiation model constructed in \citet{2023MNRAS.525..246H}. 

The ejecta-CSM interaction could generate the relativistic electrons. The frequency spectrum of the relativistic electrons is complicated and can far exceed the gyration frequency. Following the scenario of gamma-ray bursts, the electron energy ($E$) satisfies a power-law distribution with the index of $p$ as $dN/dE = N_0 E^{-p}$, where $N$ and $N_0$ are the number density of the electrons and a scaling parameter, respectively. The synchrotron radiation from relativistic electrons with a power-law distribution also has a power-law spectrum as the emission coefficient ($j_{\nu}$) satisfies $j_{\nu} \propto \nu^{-\alpha}$, where the parameter $\alpha = (p-1)/2$. We adopt $\alpha = 1$ and $p = 3$ in this study, consistent with the previous studies.  

With the evolution of $V_{\rm sh}$ in Section~\ref{Sec301} and the assumption that a fraction of the shocked energy, $\epsilon_{\rm e}$, goes into the electron energy, the parameter $N_0$ is expressed as $N_0 = (p-2)\epsilon_{\rm e}\rho_{\rm csm}V_{\rm sh}^2E_{\rm min}^{p-2}$, where $E_{\rm min}$ is the minimum of the electron energy. With the identified electron energy distribution, the corresponding radio luminosity containing the synchrotron self-absorption effect is expressed as Equation 4 in \cite{2023MNRAS.525..246H}.

\begin{figure}[t!]
\centering
\includegraphics[width = 0.48 \textwidth]{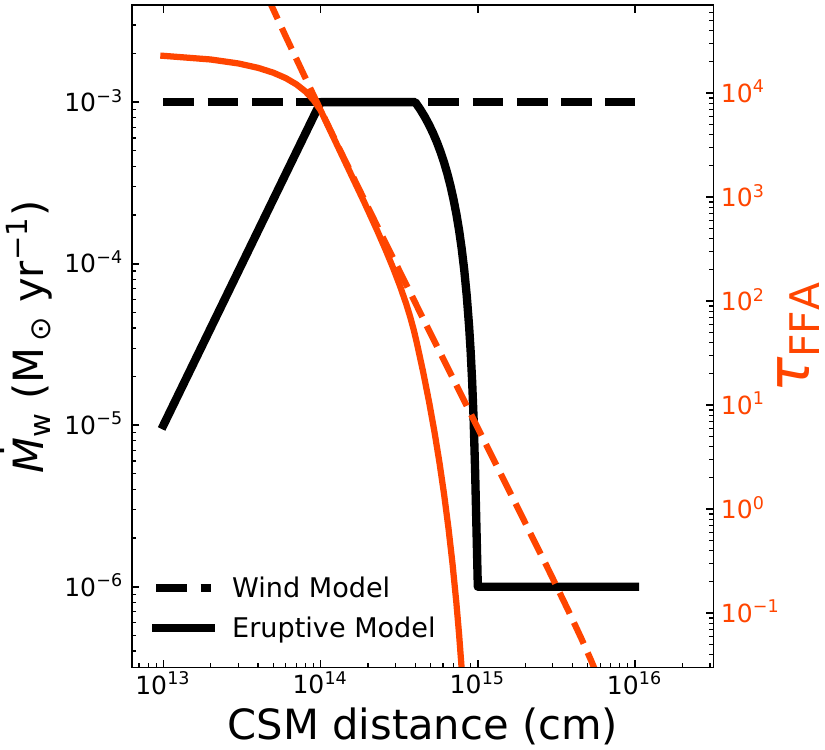}
\caption{The dashed and solid black lines are the mass-loss rate versus CSM distance for the Wind model (dashed line) and the Eruptive model (solid line), respectively. The corresponding optical depth of free-free absorption is shown in red lines.} 
\label{fig_33} 
\end{figure}

\subsection{Model updates}

Two effects are not involved in the model of \cite{2023MNRAS.525..246H}: the free-free absorption of the unshocked CSM and the fraction of nonrelativistic electrons. In the picture of \cite{2023MNRAS.525..246H}, the hypothesized CSM surrounding thermonuclear SNe is thin with the mass-loss rate of about $10^{-6}\ {\rm M}_{\odot}\,{\rm yr}^{-1}$, leading to negligible free-free opacity. Besides, a typical shock velocity in \cite{2023MNRAS.525..246H} is about $30,000\ {\rm km}\,{\rm s}^{-1}$, indicating the deviation from Equation 18 in \cite{2006ApJ...651..381C} is small, and then the nonrelativistic electron component is negligible. In our study, the mass-loss rate is at the level of about $10^{-2}\ {\rm M}_{\odot}\,{\rm yr}^{-1}$ and the shock velocity is about $10,000\ {\rm km}\,{\rm s}^{-1}$, resulting in significant free-free optical depth and a large fraction of nonrelativistic electrons. 

\subsubsection{Electron energy}

With the assumption that the electron energy goes into a power-law distribution, the average electron energy ($\bar{E}$) is expressed as $\bar{E} = \frac{p-1}{p-2}E_{\rm min}$. On the other hand, the equation below links the shock energy and $\bar{E}$,
\begin{equation}
\epsilon_{\rm e}n_{\rm i}\mu m_{\rm p}V_{\rm sh}^2/\eta = n_{\rm e}\bar{E}
\end{equation}
where $n_{\rm i}$ and $n_{\rm e}$ are the number density of ion and electron, respectively. $m_{\rm p}$ is the proton mass and $\mu$ is the mean molecular weight. $\eta$ (= 4) is a shock compression factor. After proper arrangement, $E_{\rm min}$ is expressed as below, 
\begin{equation}
E_{\rm min} = \frac{(p-2)}{(p-1)}\frac{\epsilon_{\rm e}\mu m_{\rm p}V_{\rm sh}^2}{\eta(n_{\rm e}/n_{\rm i})} 
\end{equation} 

In the previous studies, like the model of \cite{2023MNRAS.525..246H}, the author adopts the Lorentz factor ($\gamma$) to represent the electron energy as $E = \gamma m_{\rm e} c^2$, where $m_{\rm e}$ is the electron mass and $c$ is the light speed. Therefore, the value of $\gamma_{\rm min}$ is calculated by
\begin{equation} 
\label{gamma_hu23}
\gamma_{\rm min} =  \frac{(p-2)}{(p-1)}\frac{\epsilon_{\rm e}\mu m_{\rm p}V_{\rm sh}^2}{\eta(n_{\rm e}/n_{\rm i})m_{\rm e}c^2},\ {\rm and}\ \gamma_{\rm min} > 1
\end{equation} 

The caveat is that Equation~\ref{gamma_hu23} might over-estimate the electron energy and then over-estimate the parameter $N_0$, resulting in over-estimating the final radio luminosity. As a comparison, we adopt the expression $E = (\gamma - 1) m_{\rm e} c^2$ to link the Lorentz factor $\gamma$ and electron energy $E$. Hence, the actual value of $\gamma_{\rm min}$ is calculated in this paper as below,
\begin{equation} 
\label{gamma_this}
\gamma_{\rm min} =  \frac{(p-2)}{(p-1)}\frac{\epsilon_{\rm e}\mu m_{\rm p}V_{\rm sh}^2}{\eta(n_{\rm e}/n_{\rm i})m_{\rm e}c^2} + 1
\end{equation} 

Equation~\ref{gamma_this} gives a more reasonable calculation of $\gamma_{\rm min}$, $\bar{\gamma}$, and $N_0$. As shown in Figure~\ref{fig_22}, in the case of low mass-loss rate (e.g., $\dot{M}_{\rm w} = 10^{-7}\ {\rm M}_{\odot}\,{\rm yr}^{-1}$), the corresponding shock velocity is above $3\times10^{9}\ {\rm cm}\,{\rm s}^{-1}$ within about 50 days after the explosion. This relatively high shock velocity ensures the similar value of $\bar{\gamma}$ calculated either from the Equation~\ref{gamma_hu23} (adopted in \cite{2023MNRAS.525..246H}) or from the Equation~\ref{gamma_this} (adopted in this work). For the cases of higher mass-loss rate (e.g., $\dot{M}_{\rm w} = 10^{-3}\ {\rm M}_{\odot}\,{\rm yr}^{-1}$), Equation~\ref{gamma_hu23} significantly over-estimates the value of $\bar{\gamma}$ and then generates more luminous radiation as shown in the middle and right panels of Figure~\ref{fig_22}.

\begin{figure}[t!] 
\centering 
\includegraphics[width = 0.48\textwidth]{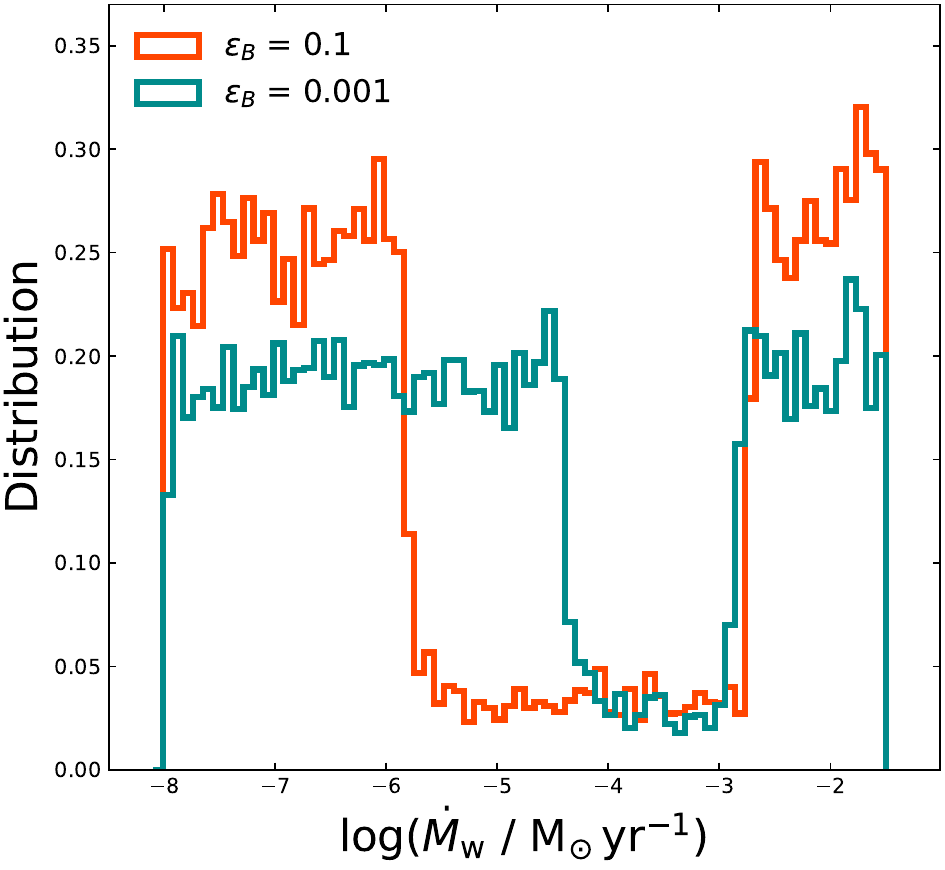}
\caption{In the context of the Wind model, this figure shows the likelihood distribution of the parameter mass-loss rate $\dot{M}_{\rm w}$ with $\epsilon_{B}$ = 0.1 (the red line) and $\epsilon_{B}$ = 0.001 (the cyan line), respectively.} 
\label{fig_mcmc} 
\end{figure}

\begin{figure*}[t!]
\centering
\includegraphics[width = 0.95 \textwidth]{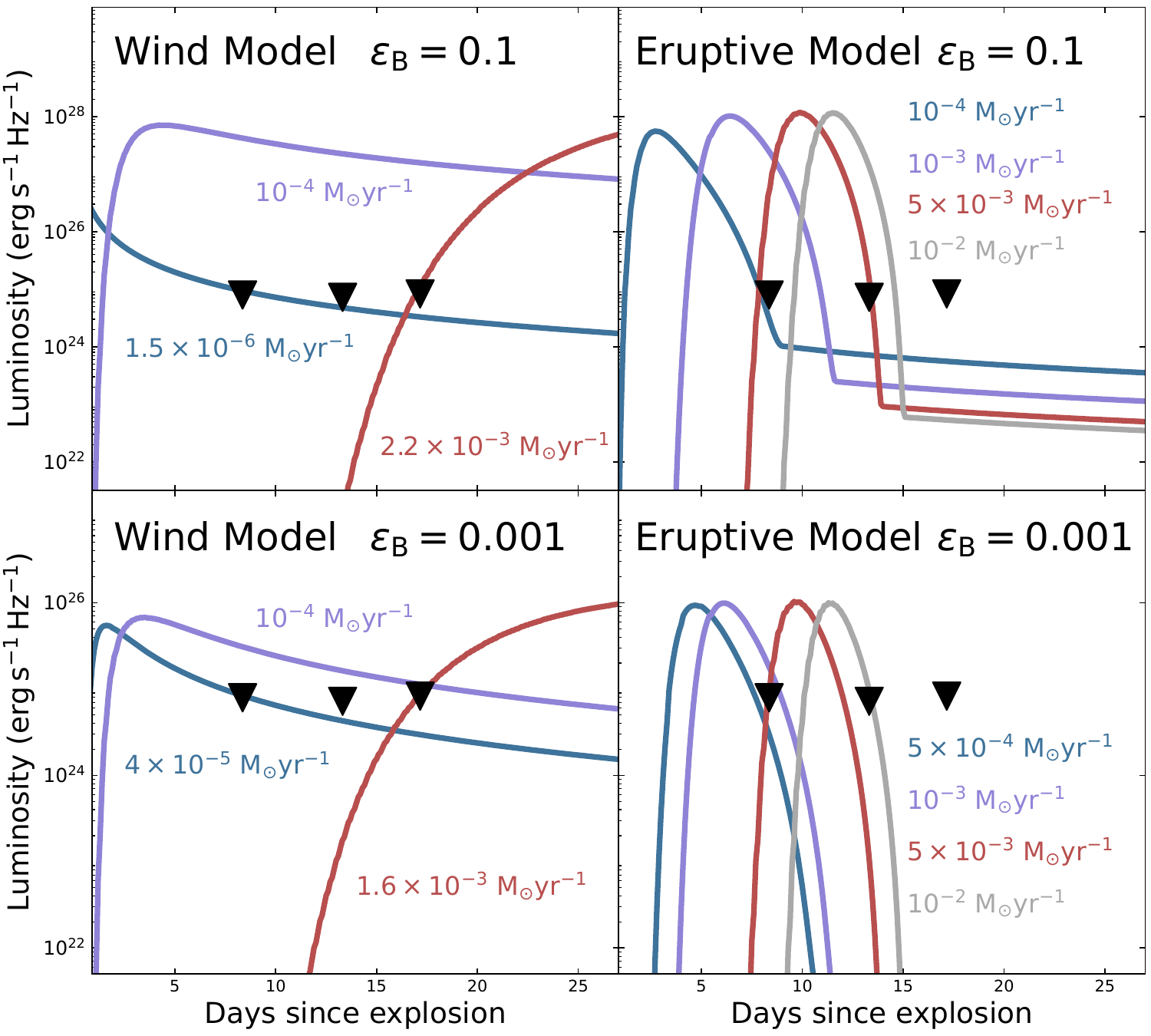}
\caption{The predicted luminosity at 230 GHz for the Wind model (left panels) and the Eruptive model (right panels) with two sets of $\epsilon_B$ as $\epsilon_B = 0.1$ (upper panels) and $\epsilon_B = 0.001$ (lower panels), respectively. The black triangles are the $3\sigma$ upper limits of SN~2024ggi obtained from ALMA observations.} 
\label{fig_44} 
\end{figure*}

\subsubsection{Free-free absorption}
The free-free absorption effect is non-negligible in this study due to the relatively high CSM density. The free-free absorption is the function of the number density of ions and electrons, frequency, and the electron temperature ($T_{\rm e}$) \citep{1975A&A....39....1P,2022ApJ...934....5Y}. The free-free optical depth and opacity are derived as 
\begin{equation} 
\tau^{\rm FFA}_{\nu} = \int \kappa^{\rm FFA}_{\nu} n_{\rm e}n_{\rm i} ds
\end{equation}
\begin{equation} 
\kappa^{\rm FFA}_{\nu} = 4.74\times10^{-27}\left(\frac{\nu}{1 \rm GHz}\right)^{-2.1}\left(\frac{T_{\rm e}}{10^5\ {\rm K}}\right)^{-1.35}
\end{equation}
where $T_{\rm e} = 5\times10^4\ {\rm K}$ adopted in our study. For simplicity, we use $\tau_{\rm FFA}$ to represent the optical depth of the free-free absorption at 230 GHz in this paper.  As shown in Figure~\ref{fig_33}, $\tau_{\rm FFA}$ of the Eruptive model rapidly decreases to be negligible within the distance of $R_3$ ($\sim 10^{15}$ cm in Figure~\ref{fig_33}), indicating that the generated millimeter-band radiation will escape outside the unshocked CSM when the shock propagates near the position of $R_3$.

\begin{figure*}[t!]
\centering
\includegraphics[width = 0.95 \textwidth]{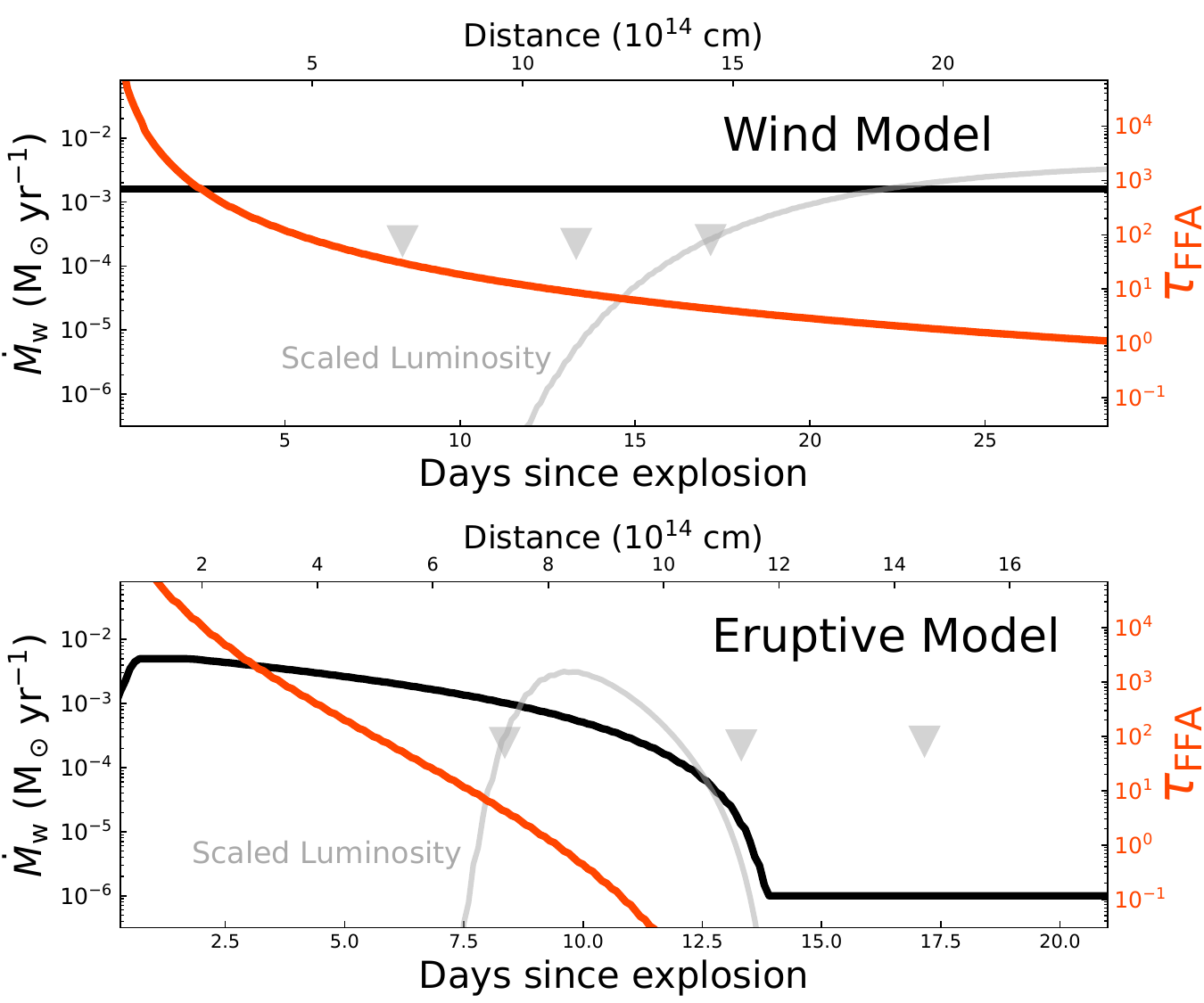}
\caption{The black lines are the mass-loss rate of the shocked CSM for the Wind model (upper panel) and the Eruptive model (lower panel), and the red lines are the corresponding optical depth of the free-free absorption. For the comparison, the gray lines are the predicted scaled luminosity at +230 GHz relating to $\epsilon_B = 0.001$ for the Wind model and the Eruptive model, and the triangles are the upper limit of SN~2024ggi. The top axis (labeled as 'Distance') is the distance of the shocked region to the SN.} 
\label{fig_55} 
\end{figure*}

\section{Results} 
\label{SecIV}

Our ALMA observations obtain high sensitivity and give a deep upper limit as $3\sigma$ RMS $<$ 0.15 mJy. The corresponding luminosity is about $8\times10^{24}\ {\rm erg}\,{\rm s}^{-1}\,{\rm Hz}^{-1}$. Such a lower luminosity limit strictly constrains the model. However, we do not apply the ejecta$-$CSM interaction to fit the early-time X-ray and optical light curves. Therefore, it is hard to constrain each parameter, like the ejecta energy and ejecta mass. For simplicity and being consistent with previous studies, we assume that $E_{\rm ej} = 1.5\times10^{51}\ {\rm erg}$ and $M_{\rm ej} = 4\ {\rm M}_{\odot}$. Besides, we consider two situations with ($\epsilon_B = 0.1$, $\epsilon_{\rm e} = 0.1$) and ($\epsilon_B = 0.001$, $\epsilon_{\rm e} = 0.1$) to represent the relatively high magnetic energy and low magnetic energy, respectively. $\epsilon_B$ is the ratio of the magnetic energy density and the thermal energy density. This low magnetic energy is also discussed in \cite{2006ApJ...641.1029C}. 

For the Wind model, the only free parameter is the mass-loss rate. As shown in Figure~\ref{fig_mcmc} and the left panels of Figure~\ref{fig_44}, we ruled out the range $\dot{M}_{\rm w} \sim 1.5\times10^{-6} - 2.2\times10^{-3}\ {\rm M}_{\odot}\,{\rm yr}^{-1}$ for $\epsilon_{\rm B} = 0.1$ and $\dot{M}_{\rm w} \sim 4\times10^{-5} - 1.6\times10^{-3}\ {\rm M}_{\odot}\,{\rm yr}^{-1}$ for $\epsilon_{\rm B} = 0.001$, respectively. The possible low value of $\dot{M}_{\rm w}$ is in contrast with the previous studies of SN~2024ggi, that the mass-loss rate is at the level of about $10^{-3} - 10^{-2}\ {\rm M}_{\odot}\,{\rm yr}^{-1}$ revealed from the early-time spectroscopic features. On the other hand, the possible high value of $\dot{M}_{\rm w}$ indicates the radial distribution of CSM could extend to a few $10^{15}$ cm (as shown in the upper panel of Figure~\ref{fig_55}), in contrast with the disappearance of the narrow H$\alpha$ emission line within a few days since the discovery. Therefore, the Wind model may not be the possible radial distribution of CSM surrounding SN~2024ggi.

For the Eruptive model, there are a few parameters to describe the distance-variant distribution of the mass-loss rate ($R_0$, $R_1$, $R_2$, $R_3$, $\dot{M}_{\rm w,0}$, and $\dot{M}_{\rm w,min}$), and it hardly to constrain these parameters with the radio data only. We adopted the same piece-wise function as Equation~\ref{eq_Mw} to match the early-time light curve of SN~2024ggi in an upcoming work \citep{Yan_inprep}, and we roughly identify that $R_0 = 10^{13}$ cm, $R_1 = 1\times10^{14}$ cm, $R_2 = 2\times10^{14}$ cm, $R_3 = 1.2\times10^{15}$ cm, and $\dot{M}_{\rm w,min} = 10^{-6}\ {\rm M}_{\odot}\,{\rm yr}^{-1}$. Hence, the only free parameter for the Eruptive model in this paper is the $\dot{M}_{\rm w,0}$. Our results indicate that $\dot{M}_{\rm w,0}$ is at the level of about $5\times10^{-3}\ {\rm M}_{\odot}\,{\rm yr}^{-1}$ for both $\epsilon_{B} = 0.1$ and $\epsilon_{B} = 0.001$, or less than $10^{-4}\ {\rm M}_{\odot}\,{\rm yr}^{-1}$ for $\epsilon_B = 0.1$ and less than $5\times10^{-4}\ {\rm M}_{\odot}\,{\rm yr}^{-1}$ for $\epsilon_B = 0.001$ as shown in the right panels of Figure~\ref{fig_44}. From the perspective of the early-time optical light curve and ionized emission lines, the CSM surrounding SN~2024ggi should be dense and close to the progenitor star. Therefore, we prefer that $\dot{M}_{\rm w,0} \sim 5\times10^{-3}\ {\rm M}_{\odot}\,{\rm yr}^{-1}$ for the Eruptive model. 

The mass-loss rate for the Eruptive model from our ALMA data aligns with the results from spectral model comparisons by \cite{2024ApJ...970L..18Z}, in which the early-time spectra of SN~2024ggi are consistent with the modeled spectra from \cite{2017A&A...603A..51D}. The model adopted by \cite{2024ApJ...970L..18Z}, which involves a mass-loss rate of $5\times10^{-3}\ {\rm M}_{\odot}\,{\rm yr}^{-1}$ and an outer radius of the dense CSM of $5\times10^{14}\ {\rm cm}\,{\rm s}^{-1}$, is similar to our results of $\dot{M}_{\rm w,0}$ and $R_2$. This consistency suggests that the Eruptive model might be the possible distribution of CSM surrounding SN~2024ggi, a potential that should be explored further. Unfortunately, the gap in the ALMA observation between +8 and +13 days after the discovery missed the opportunity to capture the millimeter-band signal relating to the ejecta-CSM interaction in the scenario of the Eruptive model. Therefore, the necessity of multi-epoch ALMA observation with the interval of a few days for the type II SNe with the early-time ionized emission lines is underscored.

\section{Discussions and Conclusions}
\label{SecV}

This paper presented the early-time millimeter-band observation of SN~2024ggi by ALMA with three epochs of $+$8, $+$13, and $+$17 days after the discovery. The $3\sigma$ image sensitivity is less than 0.15 mJy, and the corresponding luminosity is about $8\times10^{24}\ {\rm erg}\,{\rm s}^{-1}\,{\rm Hz}^{-1}$. The null detection with this deep sensitivity may indicate a large amount of shocked nonrelativistic electrons or significant free-free absorption. Therefore, we updated the radio-band radiation model shown in \cite{2023MNRAS.525..246H} by considering the fraction of nonrelativistic electrons and the free-free absorption opacity relating to the unshocked CSM along the line of sight. These two factors significantly decrease the radio luminosity in the case of SN~2024ggi due to dense CSM. 

SN~2024ggi, like SN~2023ixf and other Type II SNe with short-lived high-ionized emission lines, is surrounded by a dense CSM close to the progenitor star. As a result, we firmly favor the Eruptive model over the Wind model in describing the radial distribution of CSM surrounding SN~2024ggi. 

In the context of an Eruptive model, the radial distribution of CSM could roughly be divided into two regions: the inner high-density region and the outer low-density region. When a shock emerges from the envelope of the progenitor star, the high-energy photons ionize the circumstellar gas close to the SN, resulting in the fast-evolving emission lines. The subsequent interaction of the ejecta and the inner high-density CSM may influence the early-time light curve. When the shock propagates within the inner high-density CSM, the generated millimeter-band radiation cannot escape the unshocked CSM due to the significant free-free absorption. As the shock reaches the region close to the low-density CSM, the millimeter-band optical depth of the free-free absorption drops below about 10, allowing us to potentially detect the associated signal (as shown in the lower panel of Figure~\ref{fig_55}). 

Given the significant free-free absorption effect, the radio radiation is unable to escape from the unshocked CSM, even for the millimeter band. Therefore, sub-millimeter observations are crucial to capture the synchrotron radiation. To further refine our understanding of the distribution of mass-loss rate, we anticipate the need for multi-epoch observations at both millimeter and sub-millimeter bands of this type of target with intervals of a few days soon after the putative explosion date. This is a promising avenue for future research, as the radio signal rapidly decreases and becomes undetectable in the scenario of the Eruptive model.


\section{Acknowledgments}
This paper makes use of the following ALMA data: ADS/JAO.ALMA$\#$2023.A.00026.T. ALMA is a partnership of ESO (representing its member states), NSF (USA) and NINS (Japan), together with NRC (Canada), NSTC and ASIAA (Taiwan), and KASI (Republic of Korea), in cooperation with the Republic of Chile. The Joint ALMA Observatory is operated by ESO, AUI/NRAO and NAOJ.
We would like to thank Ji Yang and Chentao Yang for the helpful discussions on the ALMA data.
This work is supported by the National Natural Science Foundation of China (NSFC grants 12288102, 12403049, and 12033003) and the China Manned Spaced Project (CMS-CSST-2021-A12). 
Maokai Hu acknowledges the support from the Postdoctoral Fellowship Program of CPSF under Grant Number GZB20240376, and the Shuimu Tsinghua Scholar Program 2024SM118. 
Y.A. acknowledges the support from the National Natural Science Foundation of China (NSFC grants 12173089), the Natural Science Foundation of Jiangsu Province (BK20211401), and the “Light of West China” Program (No. xbzg-zdsys-202212). 
X. Wang acknowledges the support from the Tencent Xplorer Prize. 


%

\vspace{5mm}







\bibliography{sample631}{}
\bibliographystyle{aasjournal}



\end{document}